\documentclass[10pt]{iopart} 

\usepackage[latin1]{inputenc}
\usepackage{graphicx}
\usepackage[colorlinks,citecolor=blue,urlcolor=blue,linkcolor=blue]{hyperref}
\usepackage{siunitx}

\expandafter\let\csname equation*\endcsname\relax
\expandafter\let\csname endequation*\endcsname\relax

\usepackage{amsmath}
\usepackage{braket}

\usepackage{iopams}
\usepackage{widetable}
\usepackage{bm}
\usepackage{cite}
\usepackage{dsfont}
\usepackage{booktabs}
\usepackage{supertabular}
\usepackage{multirow}
\usepackage{rotating}

\begin{document}

\title[]{Uncovering dispersion properties in semiconductor waveguides to study photon-pair generation}

\author{K. Laiho$^1$,  B. Pressl$^1$,  A. Schlager$^1$, H. Suchomel$^2$, M. Kamp$^2$,  S. H\"ofling$^{2,3}$, C. Schneider$^2$ and G. Weihs$^{1}$}

\address{$^1$Institut f\"ur Experimentalphysik, Universit\"at Innsbruck, Technikerstra\ss e 25, 6020 Innsbruck, Austria}

\address{$^2$Technische Physik, Universit\"at W\"urzburg, Am Hubland,  97074 W\"urzburg, Germany}

\address{$^3$School of Physics $\&$ Astronomy, University of St Andrews, St Andrews, KY16 9SS, United~Kingdom}

\eads{\mailto{kaisa.laiho@uibk.ac.at}}

\date{\today}

\begin{abstract}
We investigate the dispersion properties of ridge Bragg-reflection waveguides to deduce their phasematching characteristics.
These are crucial for exploiting them as  sources of parametric down-conversion (PDC). In order to estimate the phasematching bandwidth we first determine the group refractive indices of the interacting modes via Fabry-Perot experiments in two distant wavelength regions.  Second, by measuring the spectra of the emitted PDC photons we gain access to their group index dispersion. Our results offer a simple approach for determining the PDC process parameters in the spectral  domain and provide an important feedback for designing such sources, especially in the broadband case.\\

\end{abstract}



\maketitle

\ioptwocol

\section{Introduction}
\label{sec:intro}

Non-linear optics provides a variety of  different frequency conversion processes for implementing diverse photonic states for quantum informatics. Especially intriguing is the process of parametric down-conversion (PDC) known for producing photon-number correlated twin beams at few photon level.  Due to their rich spectral structure only very few of these sources intrinsically emit light into a single optical mode \cite{P.J.Mosley2008, A.Eckstein2010}. Yet, despite the spectral multimodeness, many of the conventional PDC sources are still capable of producing indistinguishable photon pairs \cite{C.K.Hong1987}. Otherwise  their output needs to undergo further state manipulation, such as filtering, before being applicable to quantum optics tasks causing an inevitable decrease in brightness.

The spectral properties of photon pairs created in the PDC process, which obeys energy and momentum conservation, depend both on the pump laser spectrum and on  the phasematching condition in the second order non-linear optical material. Together they result in a joint spectral distribution (JSD) of the twin beams---typically called signal and idler \cite{W.P.Grice1997, T.E.Keller1997}. The standard way to measure the JSD is to scan over the signal and idler frequencies while detecting the coincident photons  \cite{W.Wasilewski2006, Poh2007,  Kuzucu2008, Avenhaus2009a, J.Chen2009, Gerrits2011}. This is usually either time consuming or takes a lot of experimental effort.  It is possible to reduce the required resources  via compressed sensing as demonstrated by observing the PDC impulse correlations \cite{Howl2013}. Recently also seeded PDC \cite{Eckstein2014, Fan2014} has been utilized to gain faster and better imaging of the JSD. In this case  the seed beams determine the spectral resolution. Also sum-frequency generation  provides a direct mapping of the spectral correlations \cite{M.Karpinski2009}.

Rastering the JSD with good resolution becomes a big effort if  signal and idler span several tens of nanometers \cite{Brida2009, Spasibko2012}.  The characteristics of the joint spectrum are mostly governed by the phasematching between the interacting modes that stems from their group indices. Luckily, the group index  can be accessed even in highly multimodal PDC waveguides such as Bragg-reflection waveguides (BRWs) via Fabry-Perot oscillations \cite{Hofstetter1997, Notomi2001, Bijlani2013, B.Pressl2015}. Thus, the spectral characteristics of the PDC emission can in principle be predicted even without non-linear interaction using \emph{linear optical experiments} only.

BRWs based on AlGaAs have become appealing for producing spectrally broadband and indistinguishable twin photons \cite{Horn2012, Guenthner2014} and for generating entanglement  \cite{Horn2013, Autebert2016} using modal phasematching between the co-propagating signal, idler and pump modes. In other words, at least one of the interacting modes is a higher order mode \cite{Helmy2011, Valles2013}. Even though accurate measurements of the refractive index of bulk AlGaAs exist \cite{Gehrsitz2000}, the effective indices of different spatial modes propagating in BRWs can usually be investigated only numerically \cite{Zhukovsky2012, Abolghasem2012, Kang2014}. While the refractive index can be simulated quite accurately, the group index calculated from these models can be far off from the true experimental value. This makes accurate predictions of the JSD very challenging.

Recently, we used Fabry-Perot measurements in BRWs to accurately measure the group index of the pump \cite{B.Pressl2015}. Here, we apply this method also in the down-conversion wavelengths to obtain the phasematching bandwidth of our BRWs. The precision of the used method depends only on the spectral properties of the measured Fabry-Perot fringes. Moreover, we record the  spectra of the individual twin beams at the few-photon level to further gain information about their group index dispersion. Our results provide  straightforward access to the spectral PDC process parameters and are useful for optimizing  the  performance of such integrated optics devices.

%
\section{Theoretical background: spectral characteristics of PDC emission}
\label{sec:theory}
The PDC twin beams, signal ($s$) and idler ($i$), are created as a squeezed vacuum, which can be approximated at low gains  as \cite{W.P.Grice1997}
\begin{equation}
\label{eq:state}
\ket{\Psi} \sim \ket{0} + \varsigma  \hspace{-1ex} \int \hspace{-1ex}  d\omega_{s} \hspace{-1ex} \int \hspace{-1ex} d \omega_{i}  f(\omega_{s}, \omega_{i})  \hat{a}^{\dagger}_{s}(\omega_{s})\hat{a}^{\dagger}_{i}(\omega_{i}) \ket{0},
\end{equation}  
where $|\mathcal{\varsigma}|^{2} \ll 1$ is the photon-pair creation probability dependent on the pumping strength and on the effective nonlinearity of the waveguide, $\hat{a}^{\dagger}_{s}$ ($\hat{a}^{\dagger}_{i}$) accounts for the photon creation at the frequency $\omega_{s}$ ($\omega_{i}$) and 
\begin{equation}
f(\omega_{s}, \omega_{i}) =\frac{1}{\mathcal{\sqrt{N}}}\
\alpha(\omega_{p}) \ 
\textrm{sinc} \bigg ( \frac{L}{2} \Delta k(\omega_{s}, \omega_{i}) \bigg)e^{-i\frac{L}{2} \Delta k(\omega_{s}, \omega_{i})}
\label{eq:JSD} 
\end{equation}
describes the joint spectral amplitude of  the PDC process taking place in an ideal non-linear optical waveguide of length $L$ and is normalized via $\mathcal{N}$ to $\int \hspace{-0.4ex}  d\omega_{s} \hspace{-0.4ex} \int \hspace{-0.4ex} d \omega_{i}  |f(\omega_{s}, \omega_{i})|^{2} = 1$. 
The pump amplitude $\alpha(\omega_{p}) = \exp(-(\omega_{p}-\omega^{c}_{p})^{2} / \sigma^{2}_{p})$ in equation \eref{eq:JSD} is usually described  in terms of the pump frequency $\omega_{p}$ by a Gaussian function with $\sigma_{p}$ being its  bandwidth  and $ \omega^{c}_{p}$ its central frequency. Due to strict energy conservation between signal, idler and pump $\omega_{p} = \omega_{s} +\omega_{i}$. Therefore, it is useful to express the frequencies $\omega_{\mu}= \omega^{0}_{\mu}+\nu_{\nu}$ in terms of detunings $\nu_{\mu}$ ($\mu = s,i,p$) from a  phasematched frequency triplet $\omega^{0}_{p} = \omega^{0}_{s} +\omega^ {0}_{i}$ and rewrite the pump amplitude  as
\begin{equation}
\alpha(\omega_{p} = \nu_{p}+\omega^{0}_{p} =  \nu_{s} +\nu_{i} +\omega^{0}_{p}) =  e^{-\frac{(\nu_{s}+\nu_{i}-\Delta)^{2}}{\sigma^ {2}_{p}}},
\label{eq:alpha}
\end{equation}
in which $\Delta = \omega^{c}_{p} -\omega^{0}_{p}$ is the detuning of the pump frequency from the phasematching.

Since in waveguided structures discrete spatial modes appear, we calculate the phase mismatch $ \Delta k(\omega_{s}, \omega_{i})$ in equation \eref{eq:JSD} for co-propagating signal, idler and pump  with the help of the effective refractive indices $n_{\mu}$ describing the mode propagation and use $k_{\mu} (\omega) = n_{\mu}(\omega) \frac{\omega}{c}$, where $c$ is the  speed of light. In the vicinity of the phasematched point, where $\Delta k^{0}= k_{s}(\omega^{0}_{s}) + k_{i}(\omega^{0}_{i}) -k_{p}(\omega^{0}_{p}) = 0$, we perform a second order series expansion of the phase mismatch by \cite{A.B.U'ren2005}
\begin{align}
\Delta  k(\omega_{s}, \omega_{i}) =&  \hspace{1ex} k_{s}(\omega_{s})+ k_{i}( \omega_{i}) - k_{p}(\omega_{p})  \\
\approx &\hspace{1ex} \Delta k^{0}  +k_{s} ^{\prime}(\omega^{0}_{s})\nu_{s} +
k_{i} ^{\prime}(\omega^{0}_{i})\nu_{i} -k_{p}^{\prime}(\omega^{0}_{p} )\nu_{p} \nonumber \\
&+k_{s} ^{\prime \prime}(\omega^{0}_{s})\frac{\nu^{2}_{s}}{2} + k_{i} ^{\prime \prime}(\omega^{0}_{i})\frac{\nu^{2}_{i}}{2}-k_{p}^{\prime \prime}(\omega^{0}_{p})\frac{\nu_{p} ^{2}}{2}, \nonumber
\label{eq:PM}
\end{align}
in which $ k_{\mu}^{\prime}(\omega) = n_{\mu}^{g}(\omega)/c $ is related to the  group index of the mode given by $n^{g}_{\mu} (\omega) = n_{\mu} (\omega) + \omega \frac{d n_ {\mu} (\omega)}{ d\omega}$ and $ k_{\mu}^{\prime\prime}(\omega) =\frac{1}{c} \frac{d n_{\mu}^{g}(\omega)}{ d\omega}$ is determined by its dispersion. 
The more compact expression 
\begin{align}
\Delta  k &=  \kappa_{s}\nu_{s} +\kappa_{i}\nu_{i} \\
+& \frac{1}{2}(K_{s}  - K_{p})\nu_{s}^{2} + \frac{1}{2}(K_{i} - K_{p})\nu^{2}_{i} - K_{p} \nu_{s}\nu_{i} \nonumber
\end{align}
with $\kappa_s = k_{s} ^{\prime}(\omega^{0}_{s})- k_{p} ^{\prime}(\omega^{0}_{p})$, $\kappa_i = k_{i} ^{\prime}(\omega^{0}_{i})- k_{p} ^{\prime}(\omega^{0}_{p})$, $K_{s} =  k_{s} ^{\prime\prime}(\omega^{0}_{s})$, $K_{i} = k_{i} ^{\prime\prime}(\omega^{0}_{i})$ and $K_{p} = k_{p} ^{\prime\prime}(\omega^{0}_{p})$
is useful for extracting the slope of the phasematching in signal and idler frequency space
by following  a constant phasematching contour via
\begin{align}
& \frac{\partial \Delta k}{\partial \nu_{i}}\Delta \nu_{i}+\frac{\partial \Delta k}{\partial \nu_{s}} \Delta \nu_{s}= 0.
\end{align}
This results in the slope or so-called phasematching tilt of
\begin{align}
\label{eq:dsdi}
& \frac{\Delta \nu_{i}}{\Delta \nu_{s}} =  -\frac{\kappa_s + K_{s} \nu_{s} - K_{p} \nu_{p}}{\kappa_i + K_{i} \nu_{i} - K_{p}\nu_{p}}, \
\end{align}
which represents the ratio  of the group index differences from twin beams to pump and is given in more general form by $ -[n^{g}_{s}(\omega^{0}_{s}  \hspace{-0.4ex} +  \hspace{-0.4ex} \nu_{s}) -n^{g}_{p}(\omega^{0}_{p}  \hspace{-0.4ex} + \hspace{-0.4ex} \nu_{p})]/[n^{g}_{i}(\omega^{0}_{i}  \hspace{-0.4ex} +  \hspace{-0.4ex} \nu_{i}) -n^{g}_{p}(\omega^{0}_{p}  \hspace{-0.4ex} +  \hspace{-0.4ex} \nu_{p})]$.

\subsection{Phasematching in the linear dispersion regime}

In order to estimate the spectral characteristics of the PDC emission, it is often adequate to consider only the first order terms in equation \eref{eq:PM}. This is typically done in the vicinity of $\Delta \approx 0$. Further,
as we are only interested in the full width at half maximum (FWHM) of the phasematching,  we replace the sinc-function in equation \eref{eq:JSD} with a Gaussian approximation by adapting their bandwidths via a constant factor $\gamma$  
\cite{W.P.Grice2001, A.B.U'ren2005}. Thus, we estimate that $\textrm{sinc} [ \Delta k(\omega_{s}, \omega_{i})L/2] \approx e^{-\gamma[ \Delta k(\omega_{s}, \omega_{i})L/2]^{2}}$. Putting equations \eref{eq:alpha} and \eref{eq:PM} into equation \eref{eq:JSD} we re-express it as 2D-Gaussian in the  form  \cite{W.P.Grice2001}
\begin{align}
\label{eq:ellipse}
&|f(\omega_{s} = \omega^{0}_{s}  \hspace{-0.5ex} + \hspace{-0.5ex}  \nu_{s}, \omega_{i} =\omega^{0}_{i}  \hspace{-0.5ex} +  \hspace{-0.5ex} \nu_{i})| \propto \\
&\exp(-\left [  \nu_{s} \ \nu_{i}  \right]
\hspace{-0.7ex} \left [ \begin{array}{cc}
\frac{1}{\sigma_{p}^{2}} +\gamma\frac{L^{2}}{4}\kappa^{2}_{s}  \hspace{-0.2ex}& \frac{1}{\sigma_{p}^{2}} +\gamma\frac{L^{2}}{4}\kappa_{s}\kappa_{i}\\
 \frac{1}{\sigma_{p}^{2}} +\gamma\frac{L^{2}}{4}\kappa_{s}\kappa_{i} \hspace{-0.2ex} & \frac{1}{\sigma_{p}^{2}} +\gamma\frac{L^{2}}{4}\kappa^{2}_{i} \\
\end{array} \right ]\hspace{-1ex}
\left[ \begin{array}{c} \nu_{s} \\ \nu_{i} \end{array} \right]). \nonumber
\end{align}
The eigenvalues of the correlation ellipse in equation \eref{eq:ellipse} are given by \cite{Laiho2009}
\begin{align}
\label{eq:sigmapm}
\frac{1}{\sigma^{2}_{\pm}} =& \frac{1}{\sigma^{2}_{p}}+\frac{\gamma L^{2}}{8}(\kappa^{2}_{s}+\kappa^{2}_{i})\\%
&\pm\sqrt{\frac{1}{\sigma^{4}_{p}}+\big(\frac{\gamma L^{2}}{8}(\kappa^{2}_{s}+\kappa^{2}_{i})\big)^{2}+\frac{1}{\sigma^{2}_{p}}\frac{\gamma L^{2}}{2}\kappa_{s}\kappa_{i}}\nonumber
\end{align}
corresponding to the minor and major semiaxes of the correlation ellipse. They  can be estimated, if the group indices of signal, idler and pump as well as the pump bandwidth and the waveguide length are known. We are only interested in the phasematching bandwidth $\sigma_{\textrm{PM}}$ that corresponds to  the minor semiaxis in equation \eref{eq:sigmapm} after letting $\sigma_{p} \rightarrow \infty$ and obtain  \cite{Laiho2009}
\begin{equation}
\frac{1}{\sigma^{2}_{\textrm{PM}}} = \gamma(\kappa^{2}_{s} +\kappa^{2}_{i})\frac{L^{2}}{4}.
\label{eq:PMsigma}
\end{equation}

\subsection{Phasematching in the curved dispersion profile}

In general, however, the group index dispersion is non-negligible causing bending of the phasematching curve so that equation \eref{eq:dsdi} does not remain constant. We investigate the case, in which the phasematching contour is probed by measuring the spectral extent of signal and idler single counts, in other words their \emph{marginal spectra}, with a continuous wave (CW) pump. 
For the signal beam this spectrum can be calculated from equation \eref{eq:state} by letting the pump spectrum $|\alpha(\omega_{p})|^{2} \rightarrow \delta(\omega_{p}-\omega^{c}_{p})$, while tracing over the idler frequency and projecting the signal on a varying frequency. If the measurement can be done with a high resolution spectrograph, the signal marginal spectrum yields
\begin{equation}
|f(\omega_{s} = \omega^{0}_{s}  \hspace{-0.4ex} +  \hspace{-0.2ex} \nu_{s})|^{2}
\propto\big | \textrm{sinc}\big(\frac{L}{2}\Delta k(\omega_{s}, \omega_{i}= \omega^{c}_{p}-\omega_{s})\big) \big|^{2}.
\end{equation}
In similar manner the idler marginal spectrum can be extracted. Thus, by recording the marginal spectra we investigate the phasemismatch $\Delta k$ in equation \eref{eq:PM} under the condition $\nu_{s}+\nu_{i} = \Delta$ and rewrite
\begin{align}
\label{eq:delta}
\Delta k
 = & \kappa_{s}\nu_{s} +\kappa_{i}\nu_{i} \\
&+ \frac{1}{2}K_{s}\nu_{s}^{2} + \frac{1}{2}K_{i}\nu^{2}_{i} - \frac{1}{2}K_{p} \Delta^{2} =  0, \nonumber
\end{align}  
from which the central frequencies of signal and idler can be solved. 
Typically the pump detuning is much smaller than the shift in signal and idler frequencies ($\Delta \ll \nu_{s,i}$). Therefore, at low pump detunings the bending of the phasematching curve  is  rather insensitive to the group index dispersion of  the pump beam and is dominated by that of signal and idler.

%
\section{Experiment}
\label{sec:sample}
\begin{figure*}
\centering
\includegraphics[width = 0.8\textwidth]{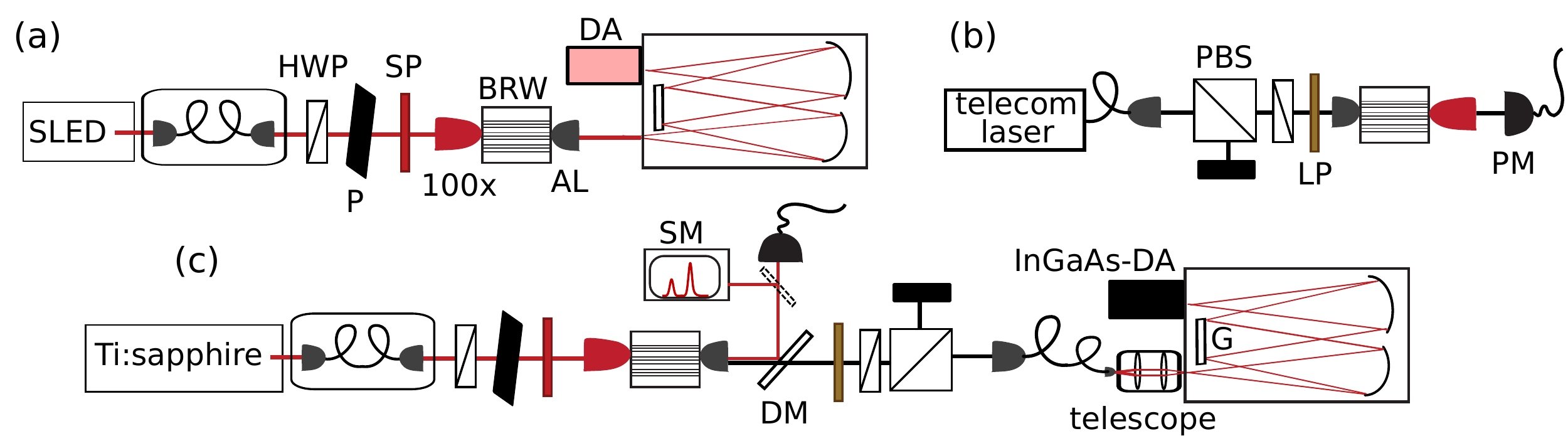}
\caption{\label{fig:exp} Experimental setups for measuring both the Fabry-Perot fringes in the (a) NIR and (b) telecommunication ranges and (c) the marginal spectra of the PDC emission. For more information and abbreviations see text.}
\end{figure*}

In our experiments we use two different BRW samples, both fabricated from the same grown wafer as in \cite{Guenthner2014, B.Pressl2015}. Ridge waveguides were defined via electron beam lithography followed by reactive ion plasma etching. The sample used in the Fabry-Perot measurements is \SI{996(4)}{\micro \meter} long and has a ridge width of \SI{6}{\micro \meter}. It is uncoated and etched down to the layer beneath the core. Meanwhile the sample employed in the measurement of marginal spectra is about \SI{1.87}{\milli\meter} long and has a ridge width of \SI{4}{\micro \meter}. It has an anti-reflection coating on the front facet for the pump wavelengths and is etched to the layer above the core.    

Both samples are phasematched for a type-II PDC process between the near infrared (NIR) pump in a higher order Bragg mode and the signal and idler modes in the  telecommunication band close to \SI{1550}{\nano\meter}. Signal and idler are cross-polarized total-internal reflection (TIR) modes---the BRW's fundamental spatial modes. According to our experience the utilization of BRWs with different ridge widths and etch depths only slightly alters the dispersion characteristics causing a shift in the phasematching wavelength that can be  about \SI{10}{\nano\meter}, nevertheless, small in relative scale. Therefore, the results gained from waveguides with the same layer structure are comparable.

Our experimental setups are shown in \fref{fig:exp}. For the Fabry-Perot measurements in the NIR  we employ in \fref{fig:exp}(a) a broadband super-luminescent light emitting diode (SLED) with central wavelength of \SI{777}{\nano\meter} and FWHM of about \SI{17}{\nano\meter} as a light source.  Its beam is mode cleaned and sent through a half-wave plate (HWP) and a sheet polarizer (P) for power and polarization control. A short pass (SP)  spectral filter guarantees that the beam is clean from undesired wavelength contributions. A 100$\times$ microscope objective couples the light to the  BRW under study and an aspheric lens with a focal length of \SI{3.1}{\milli\meter} collects the light afterwards. After that the NIR light is directed to a spectrograph with \SI{10}{\pico\meter} resolution and recorded with a detector array (DA) sensitive in the range from \SI{400}{\nano\meter} to \SI{1000}{\nano\meter} for observing the Fabry-Perot fringes as in reference \cite{B.Pressl2015}.

For the  transmission measurements in the telecommunication range we couple in \fref{fig:exp}(b) a tunable CW laser via a polarizing beam splitter (PBS) and a HWP that controls the polarization in the backward direction  to our BRW.  The Fabry-Perot fringes are measured in the spectral range from \SI{1523}{\nano\meter} to \SI{1594}{\nano\meter} with a power meter (PM). 

For measuring the marginal spectra in \fref{fig:exp}(c) we slightly modify the setup in \fref{fig:exp}(a). We now employ CW Ti:sapphire laser as a pump. After passing through the BRW, the pump is separated from the light in the telecommunication range with a dichroic mirror (DM) and sent either to a  spectrometer (SM) having a resolution of about \SI{0.3}{\nano\meter} or to the PM. The pump power coupled through the BRW  is kept constant at approximately \SI{0.6}{\milli\watt}. The PDC emission then passes a long-pass  (LP) filter, which suppresses background light below \SI{1.4}{\micro \meter},  and gets coupled in a polarization-selective manner to a grating spectrometer mounted to an InGaAs-detector array (InGaAs-DA) operating at \SI{-40}{\celsius}. A single exposure can record a spectral band of about \SI{110}{\nano\meter}. In order to cover a larger bandwidth, its grating (G) is moved with a motorized stage. The acquisitions are stitched together via calibration points recorded with a tunable narrowband CW telecom laser resulting in a  total resolution of about 1 nm per pixel. Moreover, our InGaAs-DA has a sharp cut-off  for wavelengths longer than approximately \SI{1.65}{\micro\meter}.

\section{Accessing phasematching bandwidth via group index}
\label{sec:results1}

\begin{figure}
\centering
\includegraphics[width = 0.48\textwidth]{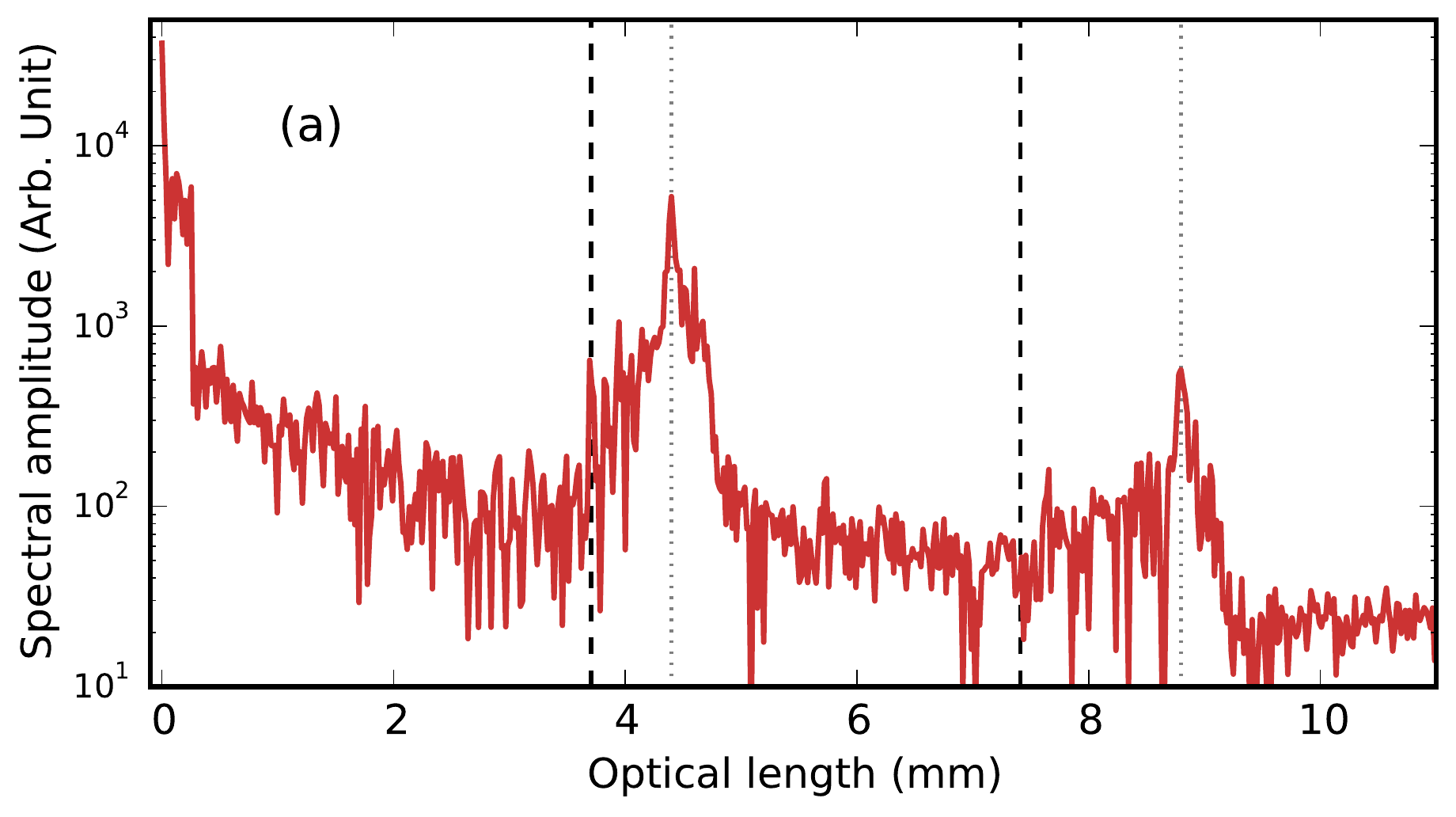}
\includegraphics[width = 0.48\textwidth]{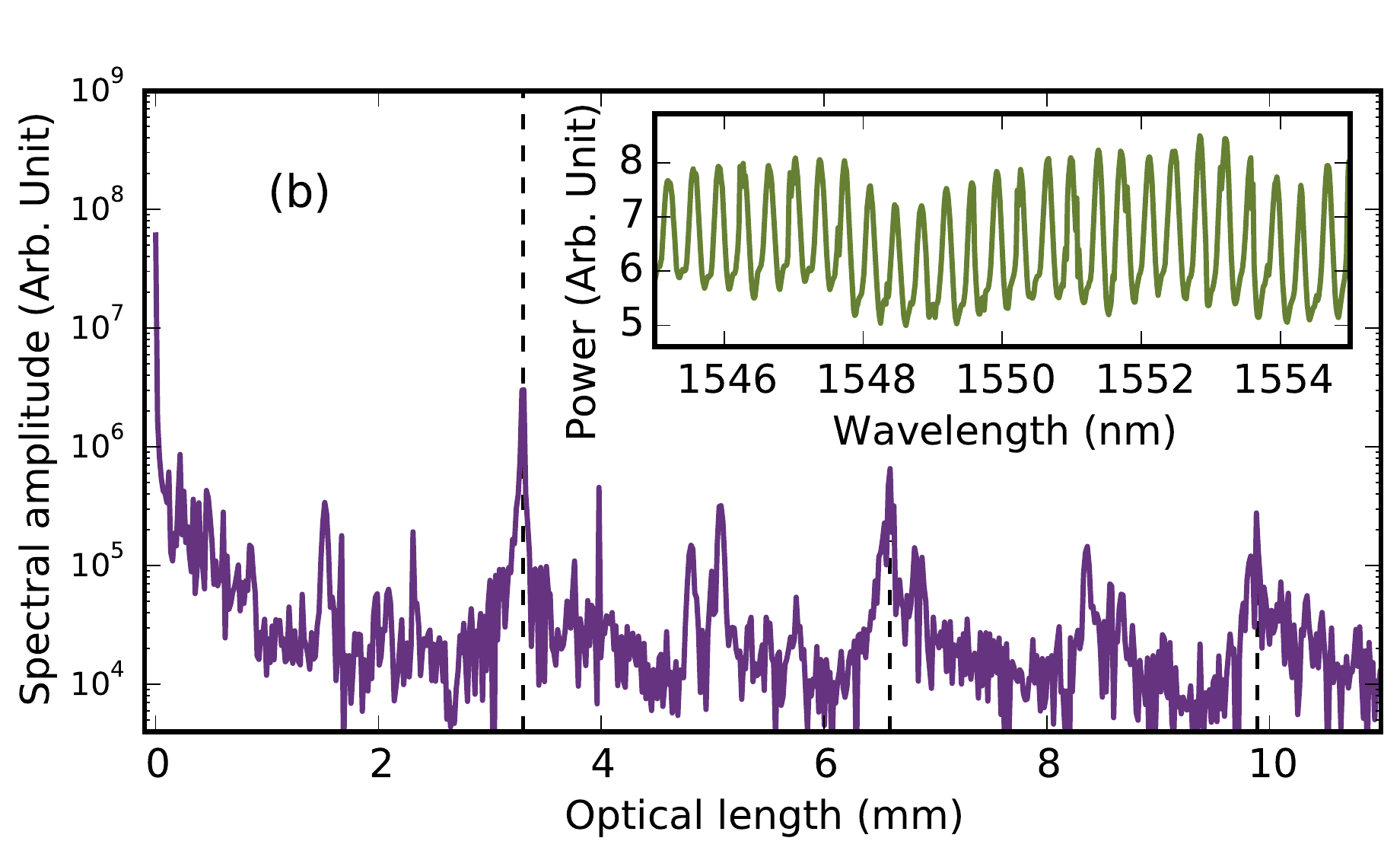}
\caption{\label{fig:fft}Fourier transformation of the measured Fabry-Perot oscillations in the (a) NIR and (b) telecommunication ranges. The inset in (b) shows a snapshot of the recorded fringes. Due to high losses in the NIR the second repeating peak is only evident for the TIR mode  and that of Bragg mode disappears in the noise floor. The dashed lines show the position of the desired modes, whereas the dotted lines indicates the TIR mode in the NIR having the group index of 4.42(3).}
\end{figure}

We start by measuring the Fabry-Perot fringes  in order to find out  the group indices of the Bragg and TIR modes and to extract the phasematching bandwidth. In figures \ref{fig:fft}(a-b) we show the Fourier transforms of the recorded fringes, while the inset in  \fref{fig:fft}(b) shows the raw data in a small wavelength range near \SI{1550}{\nano\meter}.  Repeating peaks in the Fourier domain appear at multiples of the optical length given by $n_{m}^{g}L$ in terms of the group index of the excited mode ($m$). 
As expected we see that the waveguide is highly multimodal in the NIR. The TIR mode is  most prominent and only a portion of the light is coupled to the Bragg mode, which is identified in  a way similar to that in reference~\cite{B.Pressl2015} as the mode with lowest apparent group index. In the telecommunication range practically only TIR mode is excited and the peaks appearing in between the repeating peaks  are most probably caused by other experimental imperfections such as fluctuations of the CW laser. 

Taking into account the waveguide length, we obtain $3.31(2)$ and $3.72(3)$ for the group indices of the TIR and Bragg modes at the PDC and pump wavelengths, respectively. In order to infer the phasematching bandwidth in the short BRW, we plug the measured optical lengths directly to equation \eref{eq:PMsigma}, which causes the resonator length to cancel out. Thus, the accuracy is solely  given by the number of the measured Fabry-Perot fringes and by their spectral resolution. 
Regarding the joint spectral intensity $\gamma = 0.179$ \cite{Luo2015} that adapts the FWHM of the squared sinc-function to that of squared Gaussian and we receive for the phasematching bandwidth a FWHM of \SI{3.7(3)}{\nano\meter}.
Finally, the group index difference between the PDC and pump modes divided by the speed of light is $ (-1.37\pm 0.12)$\SI{}{\pico\second/\micro\meter}, which we call $\bar{\kappa}$. We note that this is an averaged value over the measured spectral ranges. Unfortunately,  the fringe measurement is  not accurate enough to resolve the differences between signal and idler. Therefore, we next study their marginal spectra at few photon level.

%

\section{Accessing group index dispersion via marginal spectra}
\label{sec:results2}

\begin{figure*}
\centering
\includegraphics[width = 0.9\textwidth]{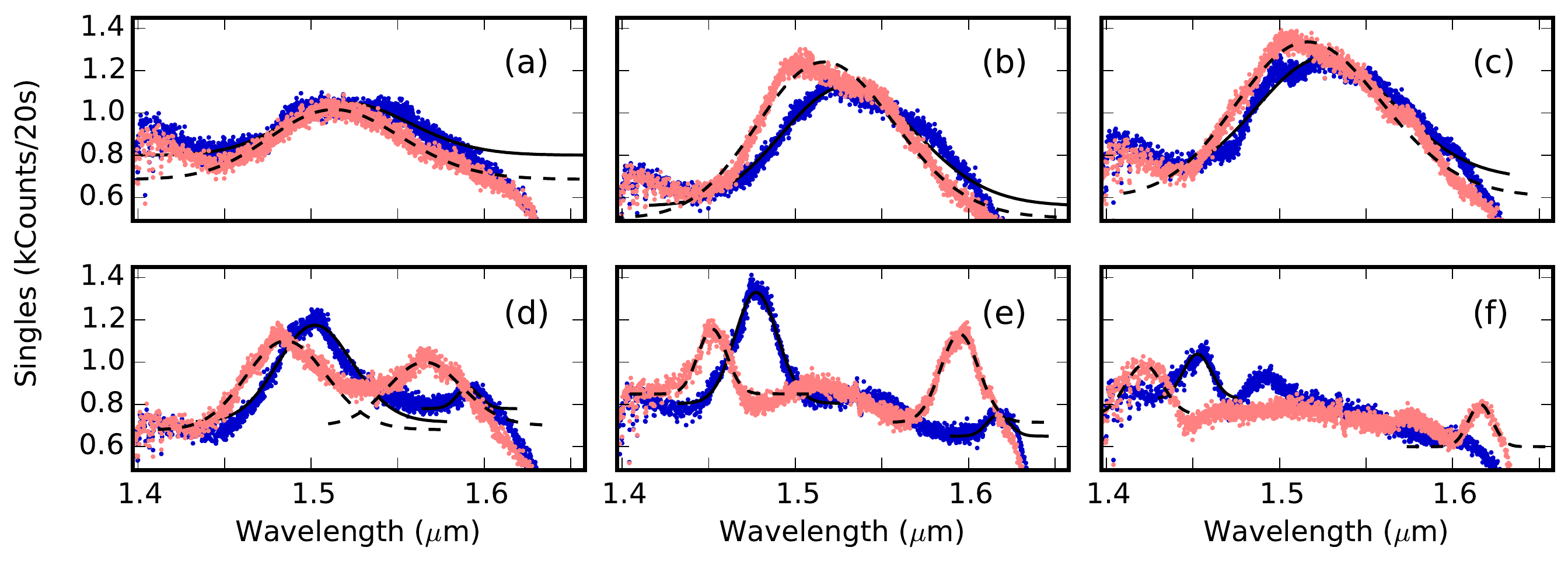}
\caption{\label{fig:marginals} Marginal spectra of signal and idler measured at the pump wavelengths (a) \SI{768.3}{\nano\meter} (b) \SI{768.0}{\nano\meter} (c)  \SI{767.8}{\nano\meter}, (d) \SI{767.5}{\nano\meter}, (e) \SI{767.1}{\nano\meter} and (d) \SI{766.7}{\nano\meter}. Light red (dark blue) symbols indicate the measured values for idler (signal), whereas the solid and dashed lines show Gaussian fits for the found  peaks. The second small peaks in (f) may be caused by laser instabilities or another phasematched process nearby.}
\end{figure*}

In order to gain information of the group index dispersion,  we record the marginal spectra of signal and idler at different pump wavelengths. Thus, we tune the pump near the degeneracy, which was observed by measuring the second-harmonic light from our BRW  at the fundamental wavelength of \SI{1535.2}{\nano\meter}.  \Fref{fig:marginals} illustrates the recorded spectra. We clearly notice that they gradually  split up into two spectral bands, i.e. have two maxima both for signal and idler. This hints to a non-negligible group index dispersion at PDC wavelengths such that equation \eref{eq:delta} has two solutions. Further, we note that there is a shift between the central wavelengths of signal and idler implicating a small difference in their group indices, which we could not resolve with our measurements in the above section \ref{sec:results1}. 

\begin{figure}
\centering
\includegraphics[width = 0.5\textwidth]{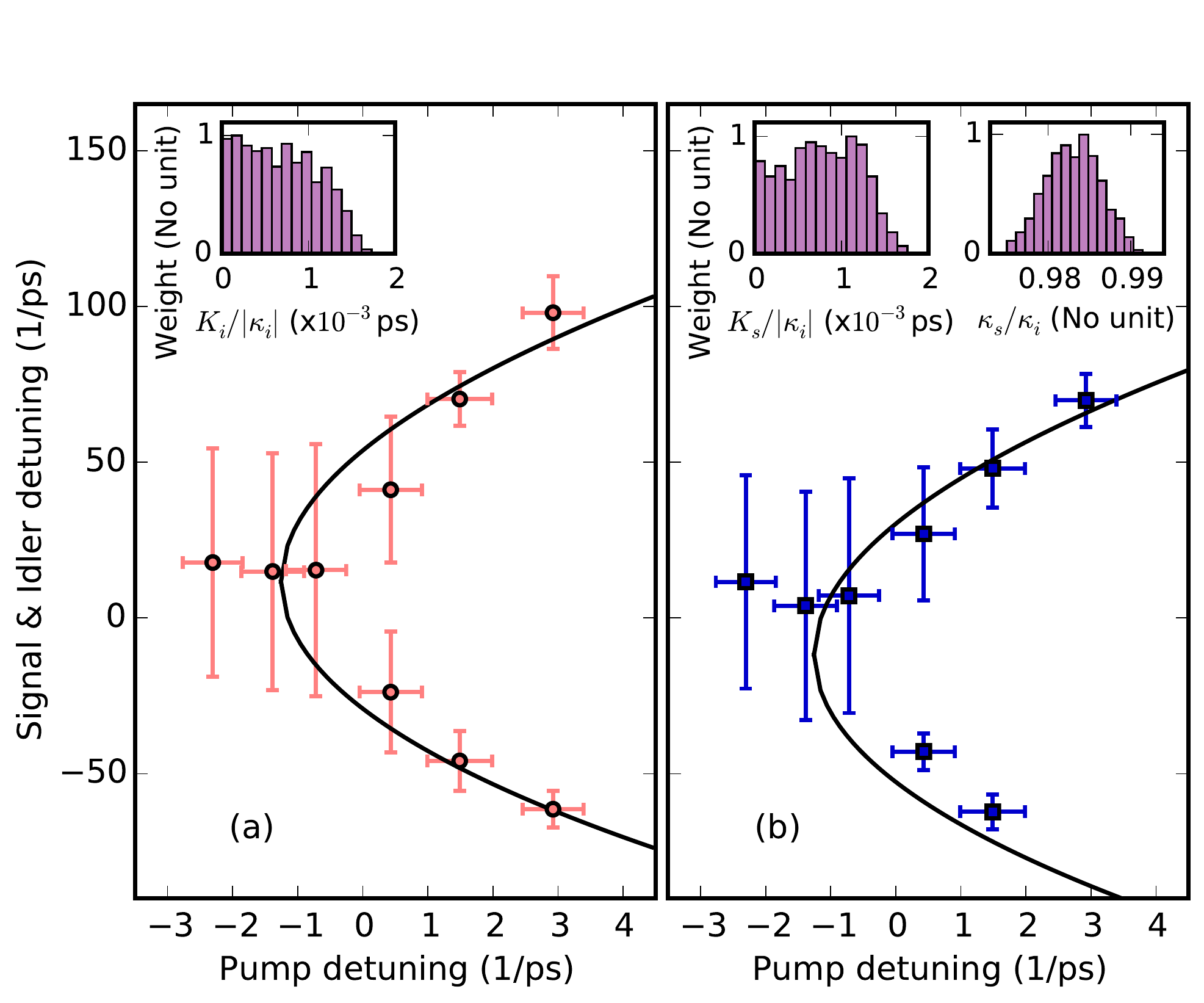}
\caption{\label{fig:phasematching} The measured frequency detunings (symbols) with respect to that of pump for (a) idler (b) signal. The vertical errorbars mark the FWHM of the measured spectra, whereas the horizontal ones indicate the spectral resolution of the bright light spectrometer. Solid lines represent the phasematching curve simulated with the parameters found with the Monte Carlo optimization. The insets show histograms for $\kappa_{s}$ and $K_{s,i}$ received from the Monte-Carlo simulation.}
\end{figure}

In \fref{fig:phasematching} we show  the fitted central frequencies for signal and idler as detunings from the degeneracy with respect to that of the pump. We see a slight offset in signal and idler detunings close to the degeneracy, which we believe is an artefact caused by  limitations in the spectral sensitivity of our InGaAs-DA. We further note that signal and idler marginal spectra  are  about \SI{90}{\nano\meter} wide near degeneracy making the accurate search of central frequencies from rather flat distributions challenging. Additionally, there is a small discrepancy between the pump detuning ($\Delta$) and that of signal ($\nu_s$) and idler ($\nu_i$), for which we ideally expect $ \Delta = \nu_{s}+\nu_{i}$.  We track this to a calibration shift between the detectors. While the InGaAs-DA is calibrated with the tunable CW laser, the pump wavelengths are measured with a separate bright light spectrometer. We match these two devices by measuring the wavelength of the second-harmonic light. However,  a small offset in pump detuning still remains.

We employ Monte Carlo simulations to find the PDC process parameters for our BRW in the spectral domain. For this purpose, we must plug in some \emph{a priori} information into equation \eref{eq:delta} since it has multiple solutions.  
First, we simulate all parameters with respect to $\kappa_{i}$, as all found solutions multiplied by a constant factor still fulfil equation \eref{eq:delta}.  We then utilize the information gained from section \ref{sec:results1} in order to receive the correct phasematching width. Second, we limit the group index of the signal to values higher than that of the idler, as expected from the measured spectra. Third, we expect that the signal and idler properties only differ slightly  from each other and that their group indices only have low dispersion over the measured spectral region. The latter can  be justified from the broadband Fabry-Perot experiment \cite{B.Pressl2015}. Finally  we investigate the phasematching properties with rather small pump detunings. In other words, while the splitting of the upper and lower spectral bands in \fref{fig:marginals} is maximally on the order of \SI{200}{\nano\meter}, that of pump is less than \SI{2}{\nano\meter}. Therefore, we disregard the term $1/2 K_{p} \Delta^{2}$  in equation \eref{eq:delta}. 

A Monte Carlo simulation with $10^{6}$ runs optimizes the parameters $\kappa_s$, $K_{s}$, $K_{i}$, while a small offset in the detuning $\Delta$ is allowed. The simulation generates these parameters from uniform distributions of large parameter ranges and accepts only those sets, for which the contour $\Delta k = 0$ lies within the vertical errorbars shown in  \fref{fig:phasematching}. We exclude the lower sideband of the signal in \fref{fig:phasematching}(b) from the fit due to the sharp cut-off in the sensitivity of our InGaAs-DA at higher wavelengths. However, we emphasize that even if only one marginal spectrum was measured, the other one  can be accurately predicted by the fitted parameters.  
The histograms gained  for $\kappa_s$, $K_{s}$ and $K_{i}$ are shown in the  insets in \fref{fig:phasematching}.  
Due to the fact that all signal detunings reside at lower frequencies than those of idler, a small group index difference of signal and  idler emerges as expected. Additionally,  sharp upper limits appear for $K_{s}$ and $K_{i}$ that are determined by the respective group index dispersions. These parameters decide, how strongly the upper and lower frequency bands in \fref{fig:phasematching} bend together with respect to the pump detuning. They also tend to opposite directions such that a large value in signal is compensated by lower one in idler, or vice versa.

\section{Phasematching in  BRWs}
\label{sec:summary}

\begin{figure}
\centering
\includegraphics[width = 0.5\textwidth]{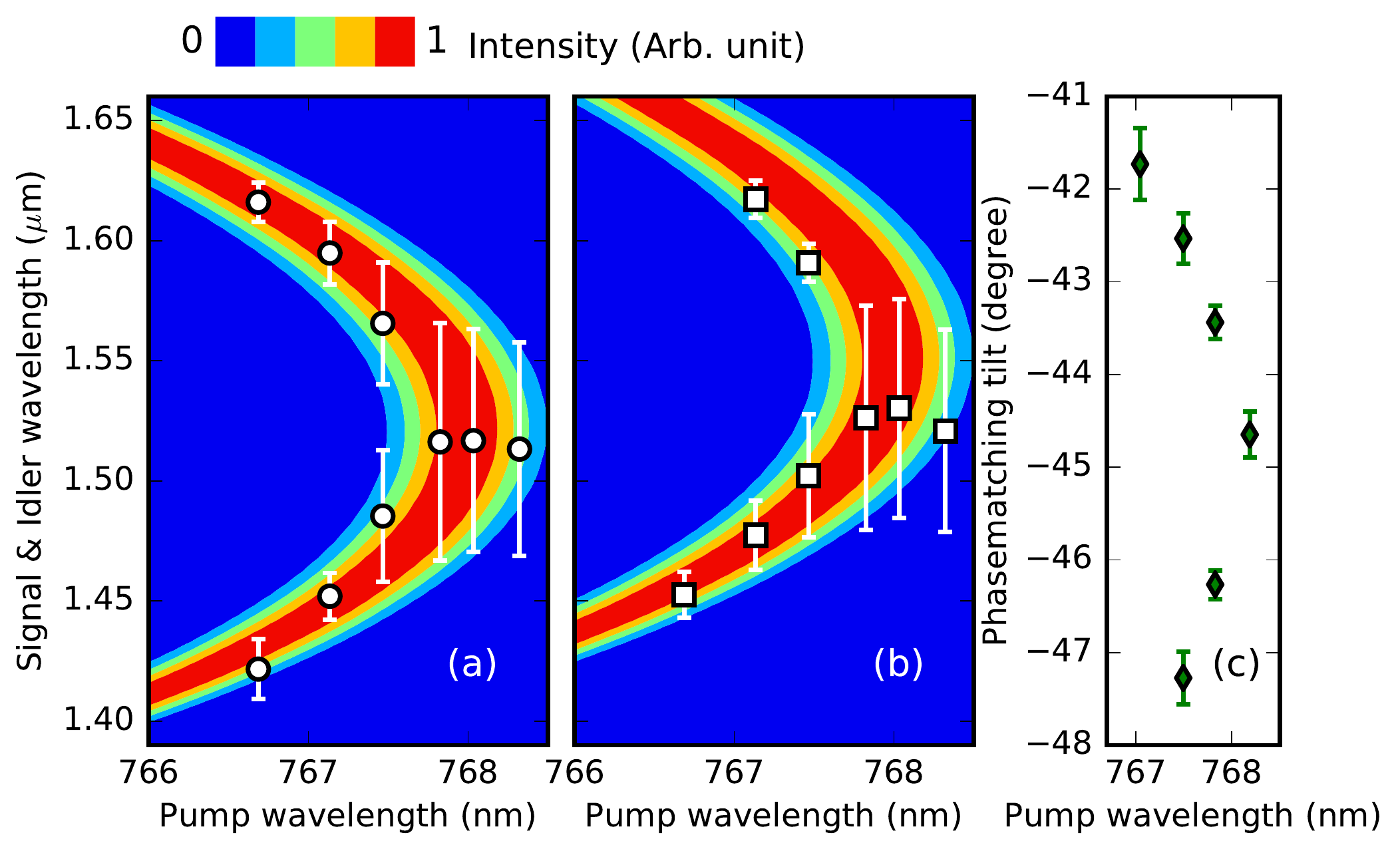}
\caption{\label{fig:PM_final} Predicted (contours) and measured (symbols) spectral properties for (a) idler and (b) signal in terms of pump wavelength together with (c) the phasematching tilt calculated by taking the arctangent from the slope in equation \eref{eq:dsdi}. The errorbars in (c) include only statistical variations  from the Monte Carlo simulation.}
\end{figure}

To summarize, we present in \tref{tab:parameters} the dispersion properties  found in sections \ref{sec:results1} and \ref{sec:results2} that govern the spectral properties of the PDC emission. By combining our results from the two experiments we can  predict the characteristics of the PDC emission. Figures \ref{fig:PM_final}(a-b) show good agreement between this prediction and the results for the marginal spectra.  We can further approximate that the group index difference between signal and idler at \SI{1550}{\nano\meter}  is about $(7.0\pm1.5)\cdot10^{-3}$. This causes a slight temporal delay between the signal and idler wavepackets when leaving BRW. Moreover,  \fref{fig:PM_final}(c) illustrates the phasematching tilt as a function of the pump wavelength for the measured central frequencies of signal and idler. It  indicates slight bending of the phasematching contours in frequency space caused by the group index dispersion.

 \begin{table}[h]
 \centering
 \begin{tabular}{clc}
 \toprule
 & Dispersion parameter & value \\
 \midrule
\multirow{5}{*}{\rotatebox[origin=c]{90}{FP}}
&$n^{g}_{\textrm{TIR}}$  at \SI{775}{\nano\meter} & 4.42(3)\\
&$n^{g}_{\textrm{Bragg}}$ at \SI{775}{\nano\meter}& 3.72(3)\\
&$n^{g}_{\textrm{TIR}}$ at \SI{1550}{\nano\meter} & 3.31(2)\\
&$\bar{\kappa}$ ($\times10^{-3}$\SI{}{\pico\second / \micro\meter})& -1.37$\pm$0.12\\
& PM bandwidth (\SI{}{\nano\meter})& 3.7$\pm$0.3\\
\midrule
 \multirow{3}{*}{\rotatebox[origin=c]{90}{MC}}
&$\kappa_s/\kappa_i$ & 0.983(4)\\
&$K_{s}/|\kappa_i|$ ($\times10^{-3}$\SI{}{\pico\second})& 0.8(5)\\
&$K_{i}/|\kappa_i|$ ($\times10^{-3}$\SI{}{\pico\second})& 0.7(5)$$\\
\bottomrule
 \end{tabular}
\caption{\label{tab:parameters}Parameters governing the spectral properties of PDC emission retrieved from the Fabry-Perot (FB) experiment in section \ref{sec:results1} and via Monte Carlo (MC) optimization for the marginal spectra in  section \ref{sec:results2}.}
 \end{table}

\section{Conclusions}
\label{sec:conclusions}

The dispersion profile of a non-linear optical material determines its phasematching characteristics and thus mainly dictates the spectral characteristics of the PDC emission. Apart from the exact phasematching wavelength given by the effective index matching of the interacting pump, signal and idler modes, the knowledge of the group refractive index and its dispersion are crucial for predicting the properties of signal and idler. We have utilized the well-known Fabry-Perot method to access the group indices at the pump and PDC wavelengths in order to estimate the phasematching bandwidth. Additionally, we have recorded spectral marginal distributions for signal and idler to infer their  group index difference and to acquire an approximation for their group index dispersion. Our results provide a straightforward method to access the spectral PDC process parameters by making only few frequency-resolved measuremements and offer means for verifying and controlling the performance of highly sophisticated, multi-layered PDC emitters.

\section*{Acknowledgements}
This work was supported by the FWF through project no. I-2065-N27, the DFG Project no. SCHN1376/2-1, the ERC project \textit{EnSeNa} (257531) and the State of Bavaria. We thank T. G\"unthner and J. Ge\ss ler  for  laboratory support and A. Wolf and S. Kuhn for assistance during sample growth and fabrication.

\appendix

\section*{References}

\begin{thebibliography}{10}

\bibitem{P.J.Mosley2008}
P.~J. Mosley, J.~S. Lundeen, B.~J. Smith, P.~Wasylczyk, A.~B. U'Ren, Ch.
  Silberhorn, and I.~A. Walmsley.
\newblock Heralded generation of ultrafast single photons in pure quantum states.
\newblock {\em Phys.~Rev.~Lett.}, 100:133601, 2008.

\bibitem{A.Eckstein2010}
A.~Eckstein, A.~Christ, P.~J. Mosley, and C.~Silberhorn.
\newblock Highly efficient single-pass source of pulsed single-mode twin beams of light.
\newblock {\em Phys. Rev. Lett.}, 106:013603, 2011.

\bibitem{C.K.Hong1987}
C.~K. Hong, Z.~Y. Ou, and L.~Mandel.
\newblock Measurement of subpicosecond time intervals between two photons by interference.
\newblock {\em Phys.~Rev.~Lett.}, 59:2044, 1987.

\bibitem{W.P.Grice1997}
W.~P. Grice and I.~A. Walmsley.
\newblock Spectral information and distinguishability in type-II down-conversion with broadband pump.
\newblock {\em Phys.~Rev.~A}, 56:1627, 1997.

\bibitem{T.E.Keller1997}
T.~E. Keller and M.~H. Rubin.
\newblock Theory of two-photon entanglement for spontaneous parametric down-conversion driven by a narrow pump pulse.
\newblock {\em Phys.~Rev.~A}, 56:1534, 1997.

\bibitem{W.Wasilewski2006}
W.~Wasilewski, P.~Wasylczyk, P.~Kolenderski, K.~Banaszek, and C.~Radzewicz.
\newblock Joint spectrum of photon pairs measured by coincidence Fourier spectroscopy.
\newblock {\em Opt.~Lett.}, 31:1130, 2006.

\bibitem{Poh2007}
H.~S. Poh, C.~Y. Lum, I.~Marcikic, A.~Lamas-Linares, and C.~Kurtsiefer.
\newblock Joint spectrum mapping of polarization entanglement in spontaneous parametric down-conversion.
\newblock {\em Phys.~Rev.~A}, 75:043816, 2007.

\bibitem{Kuzucu2008}
O.~Kuzucu, F.~N.~C. Wong, S.~Kurimura, and S.~Tovstonog.
\newblock Joint temporal density measurements for two-photon state characterization.
\newblock {\em Phys. Rev. Lett.}, 101:153602, 2008.

\bibitem{Avenhaus2009a}
M.~Avenhaus, A.~Eckstein, P.~J. Mosley, and Ch. Silberhorn.
\newblock Fiber-assisted single-photon spectrograph.
\newblock {\em Opt.~Lett.}, 34:2873, 2009.

\bibitem{J.Chen2009}
J.~Chen, A.~J. Pearlman, A.~Ling, J.~Fan, and A.~Migdall.
\newblock A versatile waveguide source of photon pairs for chip-scale quantum information processing.
\newblock {\em Opt.~Express}, 17:6727, 2009.

\bibitem{Gerrits2011}
T. Gerrits, M. J. Stevens, B. Baek, B. Calkins, A. Lita, S. Glancy, E. Knill, S. W. Nam, R. P. Mirin, R. H. Hadfield, R. S. Bennink, W. P. Grice, S. Dorenbos, T. Zijlstra, T. Klapwijk, and V. Zwiller, 
\newblock Generation of degenerate, factorizable, pulsed squeezed light at telecom wavelengths.
\newblock {\em Opt. Express}, 19:24434, 2011.

\bibitem{Howl2013}
G.~A. Howland and J.~C. Howell.
\newblock Efficient high-dimensional entanglement imaging with a compressive-sensing double-pixel camera.
\newblock {\em Phys. Rev. X}, 3:011013, 2013.

\bibitem{Eckstein2014}
A.~Eckstein, G.~Boucher, A.~Lemaitre, P.~Filloux, I.~Favero, G.~Leo, J.~E. Sipe, M.~Liscidini, and S.~Ducci.
\newblock High-resolution spectral characterization of two photon states via  classical measurements.
\newblock {\em Laser Photon. Rev.}, 8:L76, 2014.

\bibitem{Fan2014}
B.~Fan, O.~Cohen, M.~Liscidini, J.~Sipe, and V.~Lorenz.
\newblock Fast and highly resolved capture of the joint spectral density of photon pairs.
\newblock {\em Optica}, 1:281, 2014.

\bibitem{M.Karpinski2009}
M.~Karpinski, C.~Radzewicz, and K.~Banaszek.
\newblock Experimental characterization of three-wave mixing in a multimode nonlinear KTiOPO$_{4}$ waveguide.
\newblock {\em Appl.~Phys.~Lett.}, 94:181105, 2009.

\bibitem{Brida2009}
G.~Brida, V.~Caricato, M.~V. Fedorov, M.~Genovese, M.~Gramegna, and S.~P. Kulik.
\newblock Characterization of spectral entanglement of spontaneous parametric-down conversion biphotons in femtosecond pulsed regime.
\newblock {\em Europhys. Lett.}, 87:64003, 2009.

\bibitem{Spasibko2012}
K.~Yu. Spasibko, T.~Sh. Iskhakov, and M.~V. Chekhova,
\newblock Spectral properties of high-gain parametric down-conversion,
\newblock {\em Opt. Express}, 20:7507, 2012.

\bibitem{Hofstetter1997}
D.~Hofstetter and R.~L. Thornton.
\newblock Theory of loss measurements of Fabry-Perot resonators by Fourier analysis of the transmission spectrum.
\newblock {\em Opt. Lett.}, 22:1831, 1997.

\bibitem{Notomi2001}
M. Notomi,  K. Yamada,  A. Shinya, J. Takahashi, C. Takahashi, and I. Yokohama.
\newblock Extremely large group-velocity dispersion of line-defect waveguides in photonic crystal slabs.
\newblock {\em Phys. Rev. Lett.}, 87:253902, 2001.

\bibitem{Bijlani2013}
B.~J. Bijlani, P.~Abolghasem, and A.~S. Helmy.
\newblock Semiconductor optical parametric generators in isotropic semiconductor diode lasers.
\newblock {\em Appl. Phys. Lett.}, 103:091103, 2013.

\bibitem{B.Pressl2015}
B.~Pressl, T.~G\"unthner, K.~Laiho, J.~Ge\ss ler, M.~Kamp, S.~H\"ofling, C.~Schneider, and G.~Weihs.
\newblock Mode-resolved Fabry-Perot experiment in low-loss Bragg-reflection waveguides.
\newblock {\em Opt. Express}, 23:33608, 2015.

\bibitem{Horn2012}
R.~Horn, P.~Abolghasem, B.~J. Bijlani, D.~Kang, A.~S. Helmy, and G.~Weihs.
\newblock Monolithic source of photon pairs.
\newblock {\em Phys. Rev. Lett.}, 108:153605, 2012.

\bibitem{Guenthner2014}
T.~G\"unthner, B.~Pressl, K.~Laiho, J.~Ge\ss ler, S.~H\"ofling, M.~Kamp, C.~Schneider, and G.~Weihs.
\newblock Broadband indistinguishability from bright parametric downconversion in a semiconductor waveguide.
\newblock {\em J. Opt.}, 17:125201, 2015.

\bibitem{Horn2013}
R.~T. Horn, P.~Kolenderski, D.~Kang, P.~Abolghasem, C.~Scarcella, A.~Della Frera, A.~Tosi, L.~G. Helt, S.~V. Zhukovsky, J.~E. Sipe, G.~Weihs, A.~S. Helmy, and T.~Jennewein.
\newblock Inherent polarization entanglement generated from a monolithic semiconductor chip.
\newblock {\em Sci. Rep.}, 3:2314, 2013.

\bibitem{Autebert2016}
C.~Autebert, N.~Bruno, A.~Martin, A.~Lemaitre, I.~Favero C.~G.~Carbonell,  G.~Leo, H.~Zbinden, and S.~Ducci.
\newblock Integrated AlGaAs source of highly indistinguishable and energy-time entangled photons.
\newblock {\em Optica}, 3:143, 2016.

\bibitem{Helmy2011}
A.S. Helmy, P.~Abolghasem, J.~S. Aitchison, B.J. Bijlani, J.~Han, B.M. Holmes, D.C. Hutchings, U.~Younis, and S.~J. Wagner.
\newblock Recent advances in phase matching of second-order nonlinearities in monolithic semiconductor waveguides.
\newblock {\em Laser Photon. Rev.}, 5:272, 2011.

\bibitem{Valles2013}
A.~Valles, M.~Hendrych, J.~Svozilik, R.~Machulka, P.~Abolghasem, D.~Kang, B.~J. Bijlani, A.~S. Helmy, and J.~P. Torres.
\newblock Generation of polarization-entangled photon pairs in a Bragg reflection waveguide.
\newblock {\em Opt. Express}, 21:10841, 2013.

\bibitem{Gehrsitz2000}
S.~Gehrsitz, F.~K. Reinhart, C.~Gourgon, N.~Herres, A.~Vonlanthen, and H.~Sigg.
\newblock The refractive index of Al$_{x}$Ga$_{1-x}$As below the band gap: Accurate determination and empirical modeling.
\newblock {\em J. Appl. Phys.}, 87:7825, 2000.

\bibitem{Zhukovsky2012}
S.~V. Zhukovsky, L.~G. Helt, P.~Abolghasem, D.~Kang, J.~E. Sipe, and A.~S. Helmy.
\newblock Bragg reflection waveguides as integrated sources of entangled photon pairs.
\newblock {\em J. Opt. Soc. Am. B}, 29:2516, 2012.

\bibitem{Abolghasem2012}
P.~Abolghasem and A.~S. Helmy.
\newblock Single-sided Bragg reflection waveguides with multilayer core for  monolithic semiconductor parametric devices.
\newblock {\em J. Opt. Soc. Am. B}, 29:1367, 2012.

\bibitem{Kang2014}
D.~Kang, A.~Pang, Y., Zhao, and A.~S. Helmy.
\newblock Two-photon quantum state engineering in nonlinear photonic nanowires.
\newblock {\em J. Opt. Soc. Am. B}, 31:1581, 2014.

\bibitem{A.B.U'ren2005}
A.~B. U'Ren, Ch. Silberhorn, R.~Erdmann, K.~Banaszek, W.~P. Grice, I.~A. Walmsley, and M.~G. Raymer.
\newblock Generation of pure-state single-photon wavepackets by conditional preparation based on spontaneous parametric downconversion.
\newblock {\em Laser~Phys.}, 15:146, 2005.

\bibitem{W.P.Grice2001}
W.~P. Grice, A.~B. U'Ren, and I.~A. Walmsley.
\newblock Eliminating frequency and space-time correlations in multiphoton states.
\newblock {\em Phys.~Rev.~A}, 64:063815, 2001.

\bibitem{Laiho2009}
K. Laiho, K. N. Cassemiro, and Ch. Silberhorn.
\newblock Producing high fidelity single photons with optimal brightness via waveguided parametric down-conversion.
\newblock {\em Opt. Express}, 17:22825, 2009.

\bibitem{Luo2015}
K.-H. Luo, H. Herrmann, S. Krapick, B. Brecht, R. Ricken, V. Quiring, H. Suche, W. Sohler and C. Silberhorn.
\newblock Direct generation of genuine single-longitudinal-mode narrowband photon pairs.
\newblock {\em New J. Phys.}, 17:073039, 2015.

\end{thebibliography}

\end{document}